\documentclass[letterpaper, 10 pt, conference]{ieeeconf}  % Comment this line out if you need a4paper

\IEEEoverridecommandlockouts                              % This command is only needed if 
\usepackage{algorithm}
\usepackage{algpseudocode}
\usepackage{kotex}
\usepackage{xcolor}
\usepackage{booktabs}
\usepackage{multirow}
\usepackage{xspace}
\usepackage{microtype}

\usepackage{cite}
\usepackage{amsmath}
\usepackage{amssymb}
\usepackage{amsfonts}
\usepackage[final]{graphicx}
\usepackage{textcomp}
\usepackage{etoolbox}
\usepackage{comment}
\usepackage{hyperref}
\usepackage{cleveref}
\usepackage{makecell}
\usepackage{flushend}
\usepackage{subcaption}

\title{\LARGE \bf
Modeling social interaction dynamics using temporal graph networks}
\author{J. Taery Kim$^{1,2}$, Archit Naik$^{1}$, Isuru Jayarathne$^{1}$, Sehoon Ha$^{2}$, Jouh Yeong Chew$^{1}$% <-this % stops a space
\thanks{$^{1}$Honda Research Institute Japan, Wakoshi, Saitama, Japan
        }%
\thanks{$^{2}$Georgia Institute of Technology, Atlanta, Georgia, USA
        }%
}

\begin{document}

\maketitle
\thispagestyle{empty}
\pagestyle{empty}

%%%%%%%%%%%%%%%%%%%%%%%%%%%%%%%%%%%%%%%%%%%%%%%%%%%%%%%%%%%%%%%%%%%%%%%%%%%%%%%%
\begin{abstract}
  Integrating intelligent systems, such as robots, into dynamic group settings poses challenges due to the mutual influence of human behaviors and internal states. A robust representation of social interaction dynamics is essential for effective human-robot collaboration. Existing approaches often narrow their focus to facial expressions or speech, overlooking the broader context. We propose employing an adapted Temporal Graph Networks to comprehensively represent social interaction dynamics while enabling its practical implementation. Our method incorporates temporal multi-modal behavioral data including gaze interaction, voice activity and environmental context. This representation of social interaction dynamics is trained as a link prediction problem using annotated gaze interaction data. The F1-score outperformed the baseline model by 37.0\%. This improvement is consistent for a secondary task of next speaker prediction which achieves an improvement of 29.0\%. %our method accurately estimates social gaze saliency, showcasing its effectiveness. 
  Our contributions are two-fold, including a model to representing social interaction dynamics which can be used for many downstream human-robot interaction tasks like human state inference and next speaker prediction. % and a social gaze saliency model. The former holds potential for downstream tasks like next speaker prediction or human state inference, enhancing the applicability of our approach in diverse machine learning applications related to social interaction.
  More importantly, this is achieved using a more concise yet efficient message passing method, significantly reducing it from 768 to 14 elements, while outperforming the baseline model.  
\end{abstract}

%%%%%%%%%%%%%%%%%%%%%%%%%%%%%%%%%%%%%%%%%%%%%%%%%%%%%%%%%%%%%%%%%%%%%%%%%%%%%%%%
%% editing comment

\newcommand{\todo}[1]{\textcolor{blue}{{TODO: #1}}}
\newcommand{\taery}[1]{\textcolor{violet}{{Taery: #1}}}
\newcommand{\sehoon}[1]{\textcolor{red}{{Sehoon: #1}}} 
\newcommand{\wenhao}[1]{\textcolor{blue}{{Wenhao: #1}}} 
\newcommand{\greg}[1]{\textcolor{cyan}{{Greg: #1}}}

\newcommand{\newtext}[1]{#1}
\newcommand{\original}[1]{\textcolor{magenta}{Original: #1}}
\newcommand{\eqnref}[1]{Equation~(\ref{eq:#1})}
\newcommand{\figref}[1]{Figure~\ref{fig:#1}}
\renewcommand{\algref}[1]{Algorithm~\ref{alg:#1}}
\newcommand{\tabref}[1]{Table~\ref{tab:#1}}
\newcommand{\secref}[1]{Section~\ref{sec:#1}}
\newcommand{\mypara}[1]{\noindent\textbf{{#1}.}}

%% ignore text
\long\def\ignorethis#1{}

%% abbreviations
\newcommand{\etal}{{\em{et~al.}\ }}
\newcommand{\eg}{e.g.\ }
\newcommand{\ie}{i.e.\ }

%% reference shortcuts
\newcommand{\figtodo}[1]{\framebox[0.8\columnwidth]{\rule{0pt}{1in}#1}}

%\renewcommand{\eqref}[1]{Equation~(\ref{eq:#1})}

%% frequently used mathematical structures

\newcommand{\pdd}[3]{\ensuremath{\frac{\partial^2{#1}}{\partial{#2}\,\partial{#3}}}}

%% New commands for Sehoon!
\newcommand{\mat}[1]{\ensuremath{\mathbf{#1}}}
\newcommand{\set}[1]{\ensuremath{\mathcal{#1}}}

% math macros
\newcommand{\vc}[1]{\ensuremath{\mathbf{#1}}}
\newcommand{\vEndEff}{\ensuremath{\vc{d}}}
\newcommand{\vRelMove}{\ensuremath{\vc{r}}}
\newcommand{\sSet}{\ensuremath{S}}

\newcommand{\vControl}{\ensuremath{\vc{u}}}
\newcommand{\vPoint}{\ensuremath{\vc{p}}}
\newcommand{\sSpringCoef}{{\ensuremath{k_{s}}}}
\newcommand{\sDamperCoef}{{\ensuremath{k_{d}}}}
\newcommand{\vHandle}{\ensuremath{\vc{h}}}
\newcommand{\vForce}{\ensuremath{\vc{f}}}

\newcommand{\mTransChain}{\ensuremath{\vc{W}}}
\newcommand{\mRotateTrans}{\ensuremath{\vc{R}}}
\newcommand{\sJoint}{\ensuremath{q}}
\newcommand{\vJoint}{\ensuremath{\vc{q}}}
\newcommand{\mJoint}{\ensuremath{\vc{Q}}}
\newcommand{\mMass}{\ensuremath{\vc{M}}}
\newcommand{\sMass}{\ensuremath{{m}}}
\newcommand{\vGravity}{\ensuremath{\vc{g}}}
\newcommand{\vConstr}{\ensuremath{\vc{C}}}
\newcommand{\sConstr}{\ensuremath{C}}
\newcommand{\vCOM}{\ensuremath{\vc{x}}}
\newcommand{\sGeneralForce}[1]{\ensuremath{Q_{#1}}}
\newcommand{\vStateVar}{\ensuremath{\vc{y}}}
\newcommand{\vControlVar}{\ensuremath{\vc{u}}}
\newcommand{\tr}[1]{\ensuremath{\mathrm{tr}\left(#1\right)}}

%%%%%%%%%%%%%%%%%%%%%%%%%%%%%%%%%%%%%%%%%%%%%%%%%%%%%%%%%%%%%%%%%%%
%
% Here are a bunch of macros, mostly for math.
%
%%%%%%%%%%%%%%%%%%%%%%%%%%%%%%%%%%%%%%%%%%%%%%%%%%%%%%%%%%%%%%%%%%%

\renewcommand{\choose}[2]{\ensuremath{\left(\begin{array}{c} #1 \\ #2 \end{array} \right )}}

\newcommand{\gauss}[3]{\ensuremath{\mathcal{N}(#1 | #2 ; #3)}}

\newcommand{\pctab}{\hspace{0.2in}}
\newenvironment{pseudocode} {\begin{center} \begin{minipage}{\textwidth}
                             \normalsize \vspace{-2\baselineskip} \begin{tabbing}
                             \pctab \= \pctab \= \pctab \= \pctab \=
                             \pctab \= \pctab \= \pctab \= \pctab \= \\}
                            {\end{tabbing} \vspace{-2\baselineskip}
                             \end{minipage} \end{center}}
\newenvironment{items}      {\begin{list}{$\bullet$}
                              {\setlength{\partopsep}{\parskip}
                                \setlength{\parsep}{\parskip}
                                \setlength{\topsep}{0pt}
                                \setlength{\itemsep}{0pt}
                                \settowidth{\labelwidth}{$\bullet$}
                                \setlength{\labelsep}{1ex}
                                \setlength{\leftmargin}{\labelwidth}
                                \addtolength{\leftmargin}{\labelsep}
                                }
                              }
                            {\end{list}}
\newcommand{\newfun}[3]{\noindent\vspace{0pt}\fbox{\begin{minipage}{3.3truein}\vspace{#1}~ {#3}~\vspace{12pt}\end{minipage}}\vspace{#2}}

\newcommand{\key}{\textbf}
\newcommand{\fun}{\textsc}

%\def\shortcite{\def\citename##1{}\@internalcite}

% Local Variables:
% TeX-master: "paper"
% End:

\begin{figure*}[ht]
  \centering
  \begin{subfigure}[]{0.35\textwidth}
      \centering
      \includegraphics[width=\textwidth]{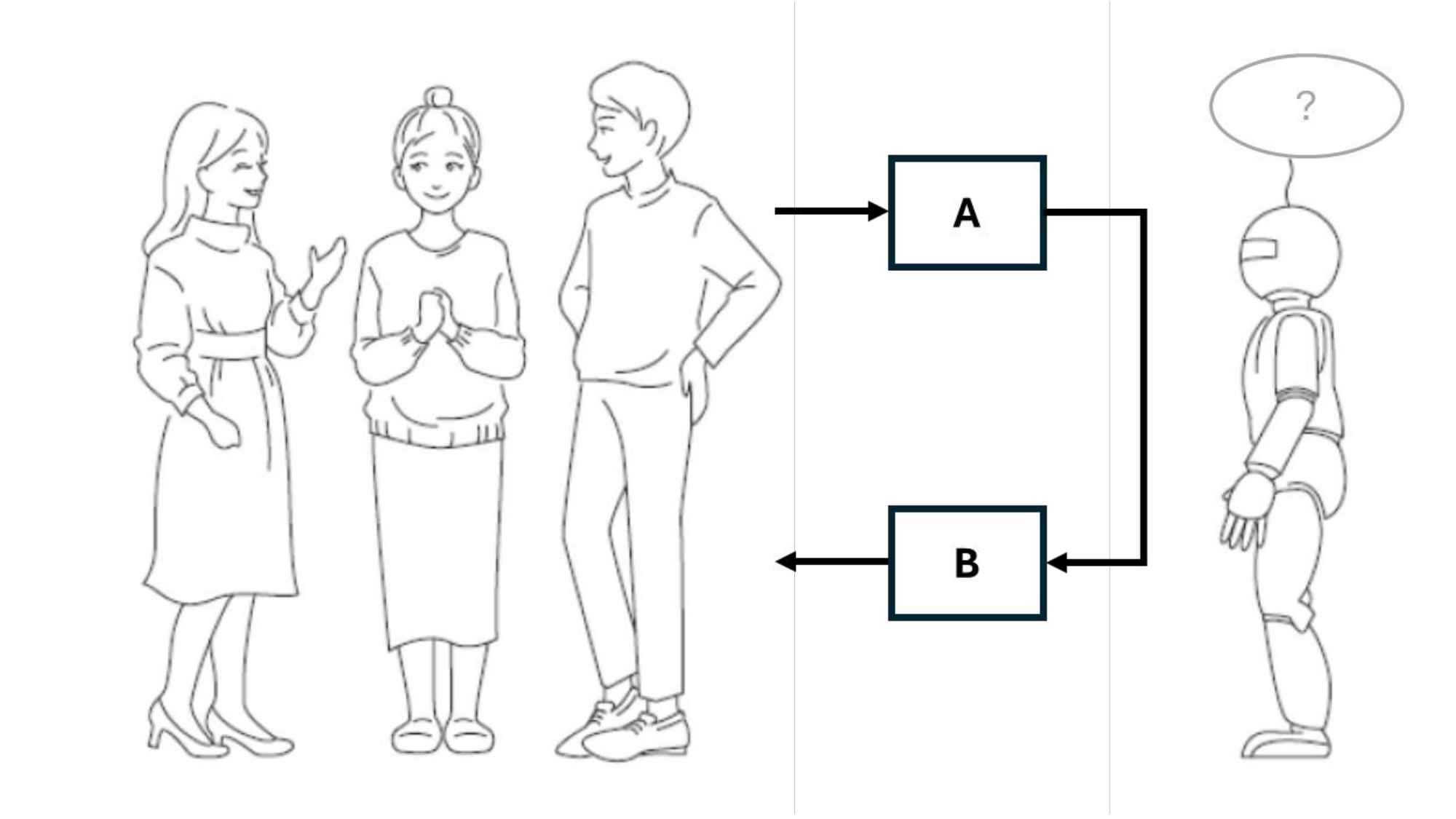}
      \caption{}
      \label{fig:teaser}
  \end{subfigure}
  \hfill
  \begin{subfigure}[]{0.6\textwidth}
    \centering
      \includegraphics[width=\textwidth]{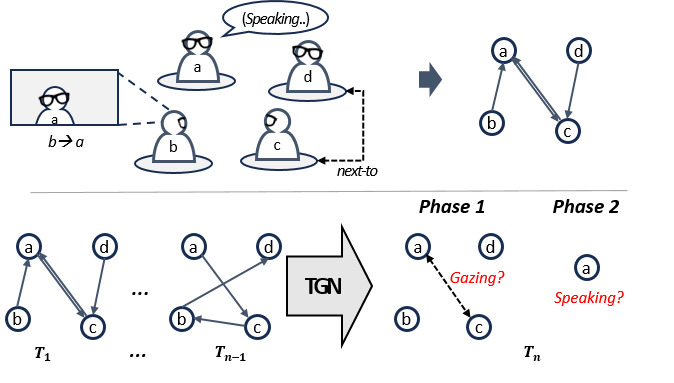}
      \caption{}
      \label{fig:graph}
  \end{subfigure}
  \caption{Overview of social interaction dynamics modeling. \textbf{(a)} Two problems are formulated for this study, A: scene perception and representation using multi-modal inputs, and B: the application of A for downstream tasks like the next speaker prediction.
  \textbf{(b)} Modeling social interaction dynamics into a temporal graph to learn via temporal graph neural networks. We tackle the next gaze prediction as problem A during phase 1 and the next speaking prediction as problem B during phase 2.
  }
  \label{fig:teaser-and-graph}
\end{figure*}

\section{Introduction}
In a hybrid society \cite{hybrid_society} where humans and intelligent systems are expected to coexist and interact with each other seamlessly, it is important for intelligent systems to understand the nuances of social interaction to realize effective collaboration. 
However, the dynamics of social interaction is complex because humans employ diverse communication strategies, often utilizing verbal communication through language accompanied by non-verbal cues such as facial expressions, gestures, and gaze interactions. It is essential to represent these multi-modal communication strategies to realize %in order to have an 
effective interaction. Non-verbal communication, known to convey a substantial portion of the message (50-70\%) \cite{holler2003iconic}, holds significant importance in our communication dynamics  \cite{burgoon2003nonverbal}. 
Numerous studies \cite{state_noc, shen2020understanding, urakami2023nonverbal} have delved into understanding these non-verbal cues. 

%The significance of eye gaze transcends that of body gestures, and other behaviors, standing out as a crucial non-verbal signal. Psychological evidence supports the notion that eyes are a cognitively special stimulus, possessing dedicated "hard-wired" pathways in the brain designed for their interpretation \cite{emery2000eyes}. 

Many studies have focused on evaluating human-robot interaction from these perspectives, such as using gaze to mediate participation imbalance between two humans and a robot \cite{gillet2022learning}, and evaluating the impact of seating position in group communication \cite{yamashita2008impact}. However, these solutions are either limited to dyadic interactions or focused on specific interaction context like gaze or seating position. In the study \cite{arslan2021speech}, a noteworthy representation of the relationship between gaze and speech is provided, yet it is confined to one-on-one interactions in an interview setting. This limits the scope of interaction to dyadic communication and does not adequately represent the scenario of multiparty interactions. Certain studies evaluated human-human \cite{joint_att,state_noc,wei2018and, nakatani2023interaction} and human-robot interaction tasks \cite{yumak2015multimodal} using multi-modalities. For example, estimation of joint attention during multi-party facilitation \cite{joint_att} and volleyball match \cite{nakatani2023interaction} scenarios achieved state-of-the-art performance. From a statistical perspective, human state tends to be related to the context of environment and body language of the other humans \cite{state_noc}. Therefore, it is important to model social interaction dynamics with a good representation of multi-modal environment context to facilitate many downstream human-robot interaction tasks, such as human state inference, next speaker prediction, and intention estimation.

However, this is not trivial because groups in real-world scenarios exhibit a dynamic and complex structure. The majority of prior studies employing deep learning methodologies often neglect this inherent structure of multiparty interactions. Instead, they heavily rely on deep learning techniques for automatic feature extraction \cite{nakatani2023interaction,joint_att}. Notably, graph neural networks (GNNs) have recently demonstrated significant potential for relational reasoning and combinatorial generalization \cite{battaglia2018relational}. In the context of multiparty interactions, representing groups as graphs has proven effective in predicting group behavior \cite{nasri2023gnn, yang2020group} or forecasting back-channeling behavior within groups \cite{sharma2022graph}. 
While the study in \cite{nasri2023gnn} seeks to leverage spatio-temporal information in group interactions, \cite{yang2020group} and \cite{sharma2022graph} overlook these informative underlying spatio-temporal patterns.
Closer to our work, a study \cite{fan2019understanding} addresses the task of nonverbal social interaction using spatio-temporal GNN. However, similar to \cite{sugano2012appearance,recasens2015they,chong2018connecting}, the study neglects multi-modal observations. Instead, it exclusively concentrates on predicting nonverbal social interaction %gaze interactions 
based on prior interactions utilizing Long Short-Term Memory.

% Having presented relevant background information showing the significance of speech \cite{arslan2021speech} and seating position \cite{yamashita2008impact} in communication dynamics, along with acknowledging the effective generalization capabilities of graphs \cite{battaglia2018relational}, the promising structured representation of groups as graphs \cite{yang2020group, sharma2022graph}, and the importance of spatiotemporal information \cite{fan2019understanding, nasri2023gnn}, we contribute the following for social gaze saliency in multiparty interactions:
The majority of preceding research concentrates on utilizing a single modality for human-robot interaction, and often overlooks the incorporation of multi-modal cues. They predominantly center around dyadic interactions and also neglect the temporal dependencies inherent in the dynamics of social interaction. Consequently, the generalization capacity of these methods to real-world scenarios is constrained.
Acknowledging these shortcomings in the existing body of work, our contributions to the field of human- robot/machine interaction is two-fold.

\begin{enumerate}
    \item representation of social interaction dynamics by integrating multi-modal features such as gaze behavior, speech, relative positions and the roles of the people and their temporal dependency. 
    \item adaptation of the Temporal Graph Networks (TGN) to enable more efficient message passing where the F1-score of our method outperformed the baseline between 29.0\% to 37.0\% for two interaction tasks. 
\end{enumerate}

%to represent social interaction dynamics by introducing integration of multi-modal features such as gaze behavior, speech, relative positions and the roles of the people and temporal dependency. We achieve this by utilizing the TGN on our dataset which captures the temporal dependency of the interaction.   

%  \textbf{1. Integration of Multi-modal Features:} We encapsulate gaze interaction within multiparty settings, leveraging multimodal cues such as speech and environmental indicators, including relative position and the roles of the involved subjects.
 
% \textbf{2. Inclusion of Temporal Dependency:}  We incorporate temporal information into multiparty gaze interaction by harnessing Temporal Graph Networks (TGN).

\section{Method}

\subsection{Problem Formulation and Graph Representation}\label{sec:method-1-problem-graph}
We formulate two problems in Fig.~\ref{fig:teaser}.  
Problem A refers to the scene perception and representation of social interaction dynamics using multi-modal inputs. 
Previous studies on human-human interaction suggested the importance of such multi-modal representation for human-robot interaction tasks \cite{multimodal_comm1,multimodal_comm2,multimodal_comm3}. 
The application of this representation for downstream tasks is defined as problem B. This paper evaluates two tasks commonly performed by intelligent systems, i.e., the next gaze task and the next speaker prediction task as problem A and B, respectively. Such abilities are crucial for intelligent systems to interact effectively with humans and to compensate for system latency.

%As suggested in previous studies \cite{emery2000eyes}, gaze interaction plays a significant role in non-verbal communication. To model gaze interactions within a group setting, w
We %employ the representation of groups 
represent social interaction as %temporal graphs 
a graph model and %try to estimate gaze interaction as 
train the model as a link prediction problem. Given graph $\text{G} = (\text{V}, \text{E}, \text{T})$ where $v\in \text{V}$ represents subjects participating in the session with a unique value from the set $\{1, \cdots, V\}$, the directed edge 
$e = (V_i, V_j) \in \text{E}$ represents that subject \textit{$V_i$} directs their gaze towards subject $V_j$ at time $\textit{t} \in \text{T}$.
We define messages passed through gaze edges with features $\textit{$f_e$} \in \text{F}$ associated with each gaze interaction.
Given the history of gaze interactions and associated features such as speaking status, seating position, and role of subjects, %we want to find 
the model estimates the probability of future gaze interaction between the humans in a group. 
% up to time \textit{t} between multiple subjects, our objective is to ascertain the probability of a gaze interaction occurring at time \textit{t+1} in the future.
\begin{figure*}[ht!] 
  \centering
  \includegraphics[width = \textwidth]{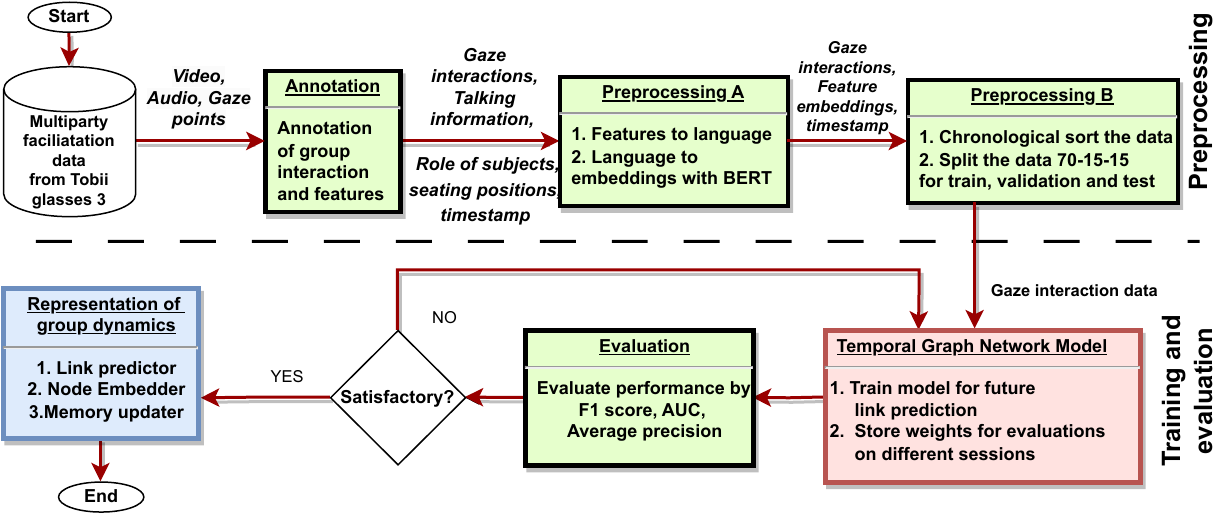}
  \caption{Preprocessing multiparty interaction data and modeling the interaction using temporal graph network model.
  % \todo{use+updated details}
  }\label{fig:figure1}
  % \Description{The figure shows the process of getting the gaze interaction and training it on Graph model to get the hidden dynamic representation of gaze interaction for multiparty discussion.}
\end{figure*}

\subsection{Multiparty Interaction Dataset}\label{sec:method-2-data}
\begin{figure}[ht] 
  \centering
  \includegraphics[width=\linewidth]{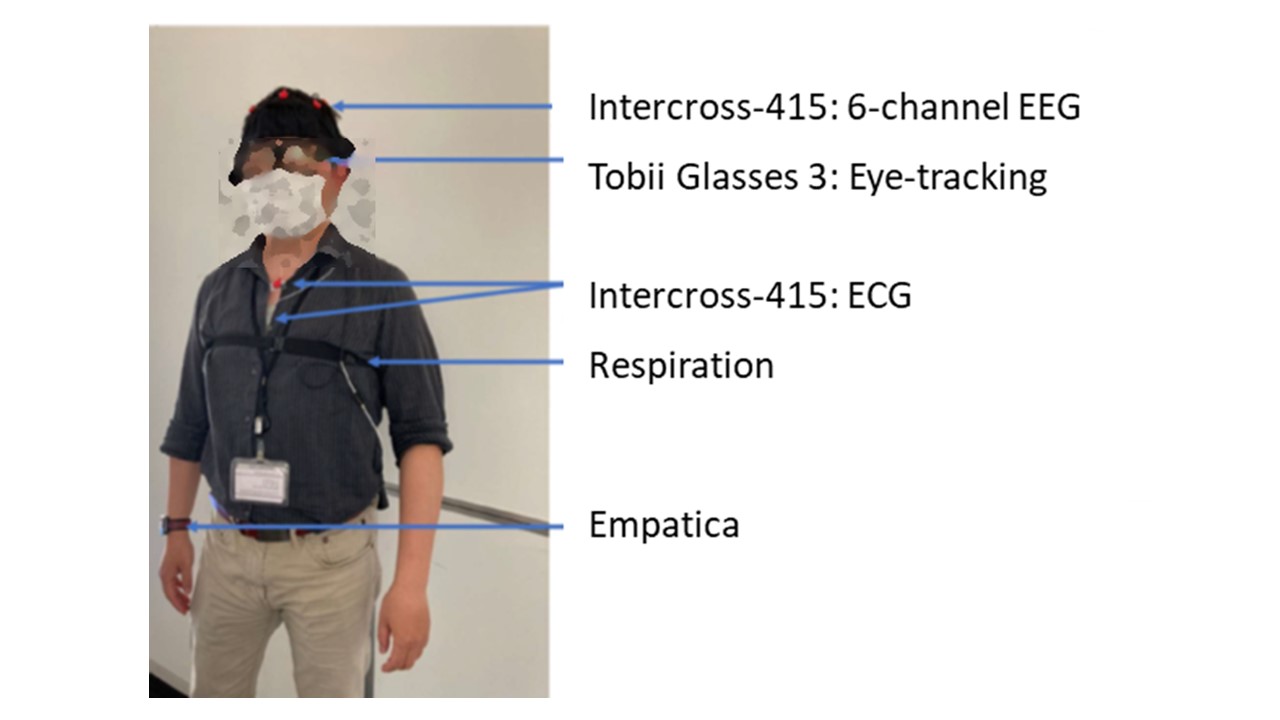}
  \caption{Sensors worn by each subject during the experiment.
  % \todo{do we need this?}
  }\label{fig:sensor}
 \end{figure}

% fumi-mpf
% \usepackage{booktabs}
% \usepackage{multirow}
\begin{table}[t]
\centering
\begin{tabular}{@{}cccccc@{}}
\toprule
\textbf{Type} & \textbf{ID} & \textbf{Facilitator type} & \textbf{\# student} & \textbf{\# facilit.} & \textbf{Total sub.} \\ \midrule
\multirow{3}{*}{D1} & S01 & None          & 5           & 0           & 5            \\
                    & S02 & None          & 5           & 0           & 5            \\
                    & S03 & None          & 4           & 0           & 4            \\ \midrule
\multirow{3}{*}{D2} & S04 & None          & 3           & 0           & 3            \\
                    & S05 & None          & 3           & 0           & 3            \\
                    & S06 & None          & 3           & 0           & 3            \\ \midrule
\multirow{3}{*}{D3} & S07 & Musician      & 3           & 1           & 4            \\
                    & S08 & Musician      & 3           & 1           & 4            \\
                    & S09 & Musician      & 3           & 1           & 4            \\ \midrule
\multirow{3}{*}{D4} & S10 & Musician      & 4           & 1           & 5            \\
                    & S11 & Musician      & 5           & 1           & 6            \\
                    & S12 & Musician      & 4           & 1           & 5            \\ \midrule
\multirow{3}{*}{D5} & S13 & Music Teacher & 5           & 1           & 6            \\
                    & S14 & Music Teacher & 4           & 1           & 5            \\
                    & S15 & Music Teacher & 5           & 1           & 6            \\ \midrule
\multirow{3}{*}{D6} & S16 & Music Teacher & 3           & 1           & 4            \\
                    & S17 & Music Teacher & 3           & 1           & 4            \\
                    & S18 & Music Teacher & 3           & 1           & 4            \\ \midrule
\multirow{3}{*}{D7} & S19 & Teacher       & 3           & 1           & 4            \\
                    & S20 & Teacher       & 3           & 1           & 4            \\
                    & S21 & Teacher       & 3           & 1           & 4            \\ \midrule
\multirow{3}{*}{D8} & S22 & Teacher       & 5           & 1           & 6            \\
                    & S23 & Teacher       & 5           & 1           & 6            \\
                    & S24 & Teacher       & 5           & 1           & 6            \\ \midrule
                    &     &               & \textbf{92} & \textbf{18} & \textbf{110} \\ \bottomrule
\end{tabular}
\caption{Session Description of FUMI-MPF dataset}
\label{tab:fumi-sessions}
\end{table}
We use the FUMI-MPF dataset, which stems from an experimental setting involving group social interactions with three types of facilitators~\cite{chew2023teach}.
The dataset comprises 24 group discussion sessions with 110 participants, categorized based on the number of students and type of facilitator as shown in Table~\ref{tab:fumi-sessions}.
Each 20-minute session focuses on music appreciation, and gaze interactions are collected at 100 Hz. Groups range in size from 3 to 6 individuals with fixed seating arrangements. 
Notably, sessions may involve the presence of a facilitator guiding the discussion, with the facilitator assuming roles such as a musician, music teacher, or teacher, alongside participating students who collectively represent various subject roles.
Fig.~\ref{fig:sensor} shows the sensors worn by each subject during the experiment, and further details on experimental conditions are explained in \cite{chew2023teach}.
% In the course of the group interaction, diverse multi-modal observation data is collected, falling into two main categories: human behavioral data, including video, audio, and gaze behavior, and human state data, such as EEG and ECG responses. However, this paper specifically concentrates on gaze interaction, speech information, and seating positions within the discussions. The data is gathered through Tobii Glasses 3 worn by each participant, which captures gaze points and speech information from the microphone. Due to preprocessing challenges, this paper experiments on 21  out of the 24 sessions
The FUMI-MPF dataset is processed to model group interaction into a graph structure, using gaze, speaking status, role, and seating position information. It is noteworthy that these interaction behavior can be obtained in real-time scenario utilizing models like gaze targeting \cite{gazetarget}, speaker diarization \cite{speakerdiari} and localization, face detection \cite{facedet} for gaze behavior, speaking status, and seating position, respectively. The information of subject's role can be obtained from speaker diarization and database.

% so that each gaze interaction at time \textit{t} between two individuals is linked to corresponding speech and context information data. Environmental data includes the seating position relative to the gazer and the roles of both individuals. Speech details, indicating who is talking at \textit{t}, are also included.  and dataset formulation are in ~\ref{data formulation}.

% \begin{table} [H]
% \caption{Overview of Sessions}
%   \label{tab:table1}
%   \begin{tabular}{p{0.2\linewidth} p{0.5\linewidth} p{0.2\linewidth}}
%     \toprule
%     Session Number&Number of gaze interactions \linebreak(respectively)&Subjects present\\
%     \midrule
%     1,2 &155754, 221330&4,5,6,7,8\\
%     3 &226647&4,5,7,8\\
%     4,5 &1163840,149506&1,2,3,11\\
%     6,7,8 &115737,74818,73643&6,7,8\\
%     9,11,13&369380,260031,281988&1,2,3,4,5,9\\
%     10&319862&1,2,3,4,9\\
%     12&265431&1,2,4,5,9\\
%     14,15,16&120469,112274,133109&6,7,8,10\\
%     17,18&258631,236868&6,7,8,9\\
%     19,20,21&338761,214107,180404&1,2,3,4,5,10\\
%     \hline
%   \bottomrule
% \end{tabular}
% \end{table}

\paragraph{Data Annotation}
We used Tobii glasses 3 eye tracking device for collecting gaze points and speech information of each subject. We matched those gaze points with faces detected in videos with InsightFace~\cite{deng2021masked} to get the gaze interactions. To recognize the voice activity of each subject, we used data from the microphone of Tobii glasses 3. The role and seating position of subjects were fixed for each session.

% \subsection{Data preprocessing} \label{sec:method-data}

\paragraph{Multiparty interaction dynamics feature extraction}
We first sort and chronologically arrange the annotated data based on the temporal sequence of interactions. Subsequently, for each gaze interaction involving subjects \textit{$V_i$} and \textit{$V_j$} at time \textit{t}, the following data, serving as features, is considered as follows:
\begin{enumerate} \label{features}
    \item Speaking status: Identification of subjects that are talking at time \textit{t}.
    \item Relative seating position: Information about the seating position of subject \textit{$V_j$} relative to subject \textit{$V_i$}.
    \item Role of subjects: Specification of the roles of subjects \textit{$V_i$} and \textit{$V_j$}.
\end{enumerate}
The extracted features are further encoded to be passed as messages in a graph for model learning (Sec.~\ref{sec:method-model-message}).

% \begin{table}[ht]
% \caption{Language representation of the environment}
% \vspace{-1em}
%   \label{tab:table1}
%   \begin{tabular}{p{0.2\linewidth} p{0.3\linewidth} p{0.4\linewidth}}
%     \toprule
%     Key Features&Sentence\linebreak(Static part)&Values\linebreak(Dynamic part)\\
%     \midrule
%     Talking \linebreak information & \underline{Talking subjects} talking.    	& Talking subjects = \{No one, \textit{$V_i$}, (\textit{$V_i$},\textit{$V_j$}),..(all possible ids)\}\\\\
%     Relative seating position&Gazing sitting \underline{position} being gazed.&position = \{opposite to, nearby, away from\}\\\\
%     Role& Gazing \underline{role(\textit{$V_i$})} and gazed \underline{role(\textit{$V_j$})}.&role(subject) = \{Student, Musician, Teacher, Music Teacher, unknown\}\\
%     \hline
%   \bottomrule
% \end{tabular}
% \end{table}

% \begin{table}[ht]
% \centering
% \caption{Language representation of the context information consisting message}
%   \label{tab:method-bert}
%   \begin{tabular}{p{0.2\linewidth} p{0.3\linewidth} p{0.3\linewidth}}
%     \toprule
%     Key Features&Possible scenario&Example sentence\\
%     \midrule
%     Speaking Status &\{No one, \textit{$V_i$}, (\textit{$V_i$},\textit{$V_j$}),...(all possible subjects)\}& \underline{\textit{$V_i$}, \textit{$V_j$}} talking\\\\
%     Seating Position&\{opposite to, nearby, away from\}&Gazing sitting \underline{nearby} being gazed.\\\\
%     Role&\{Student, Musician, Teacher, Music Teacher, unknown\} &Gazing \underline{Student} and gazed \underline{Teacher}.\\
%     \hline
%   \bottomrule
% \end{tabular}
% \end{table}

\subsection{Model} \label{model information}
% This subsection explains the approaches taken for this study. 
% Due to the constraint of chronological ordering of the data for the training TGN, we made separate graphs for each session and trained them independently. 
% For each session we had set of gaze interactions E with its associated features F among subjects V at a specific time T. 
% The model architecture, illustrated in Figure \ref{fig:figure1} depicts the experimental workflow. Following data preprocessing, as explained in Section \ref{data formulation}, we transform the session data into a temporal graph by utilizing PyTorch Geometric \cite{Fey/Lenssen/2019}.
% Subsequently, this data is input into the TGN model, where the model learns the hidden representation of gaze interactions. Leveraging both memory and representation, the model can forecast future gaze interactions. Once the model is trained on a specific session, the combinatorial generalization capabilities of graph networks \cite{battaglia2018relational}, enable the application of the learned representation from one session to others, encompassing different subjects and sizes. Further insights into generalization capabilities are presented in Section \ref{results:Performance}.

We model the social interaction dynamics by leveraging temporal graph networks (TGN)~\cite{rossi2020temporal} as shown in Fig.~\ref{fig:graph}.
As described in Sec.~\ref{sec:method-1-problem-graph}, subjects and their gaze interactions in a group are modeled into a graph structure, nodes and edges, respectively.
To capture the dynamics along with time flow, we incorporate the time information into the temporal graph for each session.
Then, the temporal graph model is trained in two phases, for problem A as the next gaze prediction and problem B as the next speaker prediction, %(), 
using messages generated from extracted features.
%We 

\subsubsection{Tasks and Learning Phases}
We train the temporal graph in two phases. 
The first phase is to learn the graph dynamics by predicting links through self-supervised learning.
Then, we take the learned graph to the next phase and further train for the node prediction as a supervised learning.
In this social dynamics modeling, the gaze prediction task is trained during first phase and the next speaker prediction task during the second phase, matched as problem A and B, respectively.

\subsubsection{Message Generation and Encoding}\label{sec:method-model-message}
While the original TGN work~\cite{rossi2020temporal} uses BERT language model~\cite{DBLP:journals/corr/abs-1810-04805} to encode the messages, we also examine one-hot encoding to represent the features in a message.
The examples of two encodings are demonstrated in Table~\ref{tab:method-message}
The effectiveness of each message encoder type is evaluated and compared in Sec.~\ref{sec:result-message} to discern the trade-offs between complexity and performance in the next gaze prediction task.

\textbf{BERT Encoding.}
BERT encoding provides flexibility in using features of a dynamic size.
For instance, when we want to note all the subjects who are currently speaking, the size of these features depends on both the number of subjects in a group and who are speaking, which could be more than one.
% Given the dynamic nature of interactions and variations in group sizes, the dimensions of features may differ. 
% To ensure the generalizability of our model across this dynamic data, it was imperative to devise a method for standardizing the data size.
To encode such dynamic-sized features to static-sized features, we initially translate the features into a linguistic format, as shown in Table~\ref{tab:method-message}.
Next, we extract a feature representation from this linguistic representation by combining sentences with all the features: speaking status, relative seating position, and role of subjects involved in a gaze.
Then, we feed the linguistic representation to a pre-trained BERT language model, resulting in a fixed-size representation.

\textbf{One-hot Encoding.}
The one-hot encoding simplifies the representation of message features by mapping each distinct state to a unique binary vector. 
Unlike BERT's context-sensitive embedding, one-hot encoding assigns a fixed, context-independent vector to each feature.
For instance, when there are subjects 2 and 4 are speaking in a gaze event, BERT encodes a sentence of ``Subject 2, Subject 4 talking.'' using the subject ID. 
On the other hand, we encode the same state using a one-hot encoder as ``$[0, 1, 1]$'',  where the first two represent the speaking status of two subjects involved in the given gaze event, and the last represents the state of the other subjects in the group.
Likewise, each subject’s speaking status, relative seating position, and role can be encoded separately and combined into a single feature vector. 
While this method is less nuanced than BERT, it offers computational efficiency and ease of interpretation, which can be beneficial for applications that do not require deep linguistic understanding or for datasets with a limited contextual scope.
\begin{table*}[ht]
\centering
\caption{Message formulation of BERT and One-hot approach}
  \label{tab:method-message}
\setlength\tabcolsep{2pt}
\begin{tabular}{@{}llll@{}}
% \toprule
\multicolumn{4}{l}{\shortstack[l]{\scriptsize \textit{Example Scenario}: Subjects 1, 2, 3, and 4 are in a group, sitting in a circle in the order of subject number. Subject 1,2,3 are students, and subject 4 is a teacher. \\ \scriptsize Subject 2 and 4 are speaking, while the other subjects are not. Current gaze from Subject 1 is to Subject 4.}} \\ \midrule
\multicolumn{1}{c}{\multirow{2}{*}{\textbf{Context Feature}}} &
  \multicolumn{1}{c}{\multirow{2}{*}{\textbf{Options}}} &
  \multicolumn{2}{c}{\textbf{Example}} \\ \cmidrule(l){3-4} 
\multicolumn{1}{c}{} &
  \multicolumn{1}{c}{} &
  \multicolumn{1}{l}{{\textbf{TGN-BERT Message}}} &
  \textbf{TGN-One-hot Message} \\ \midrule
Speaking Status &
  \{No one, \textit{$V_i$}, (\textit{$V_i$},\textit{$V_j$}),...(all   possible subjects)\} &
  \underline{\textit{$V_2$}, \textit{$V_4$}} talking. &
  $[0, 1]$ {\scriptsize (gazing subj., gazed subj.)}, $[1]$ {\scriptsize (others)} \\ \midrule
Seating Position &
  \{opposite to, nearby, away from\} &
  Gazing sitting \underline{nearby} being gazed. &
  [0, 1, 0] \\ \midrule
Role &
  \{Student, Musician, Teacher, Music Teacher, unknown\} &
  Gazing \underline{Student} and gazed \underline{Teacher}. &
  [1, 0, 0, 0], [0, 0, 1, 0] \\ \bottomrule
\end{tabular}

\end{table*}

\subsubsection{Batch Size}
The batch size has a significant effect on the temporal graph model that it contributes to the degradation of the quality of memory~\cite{rossi2020temporal}. 
With an increase in batch size, there is a corresponding increase in the number of gaze interactions encapsulated within the batch.
Consequently, this escalation results in the memory becoming relatively outdated for the final interaction within a larger batch-size scenario. 
That is, the batch size decides the time window for the prediction tasks in our application.
As we increase the size, the model learns to predict the future further, which makes the problem harder.

\subsubsection{Negative Edge Sampling}
The negative sampling methodology from TGN work~\cite{rossi2020temporal} randomly selects subjects based on the minimum and maximum values associated with the subjects. 
The resulting negative samples may encompass individuals who are not part of the session or, conversely, individuals who serve as actual gaze targets in the interaction (false-negative edges). 
While such an approach does not affect the learning critically for large datasets such as Wikipedia or Reddit~\cite{kumar2019predicting}, it does not sufficiently capture a realistic scenario of multiparty discussions with a small number of nodes.
To avoid those false-negative edges and better align with the multiparty setting, we exclusively sample subjects present only in the session while filtering out false samples which represent actual gaze.
Additionally, the sampling relies on the source node of the positive samples in the same batch.
Thus, the dataset does not capture the state when a subject is not gazing at anyone.
To estimate each node's state better, we inject such negative edges with destination to an empty node to the dataset. 
% Algorithm \ref{alg:Negative sampling} outlines the details of our proposed negative sampling strategy.     

\subsubsection{TGN Modules}
We use the GRU \cite{cho2014learning} as the Memory updater, Temporal graph attention (attn) as the embedding module for the nodes, identity (id) as the message function, and mean message aggregator. 
We follow the default configuration of the tgn-attn variant from TGN \cite{rossi2020temporal}, except for the message aggregator.
The choices on the TGN modules are made empirically, which we present the ablation study in Sec.~\ref{sec:result-tgnvariant}
% Detailed explanations of each module can be found in that paper.

\subsection{Implementation}
We use PyTorch Geometric~\cite{Fey2019pyg} for the implementation of the TGN model.
We split the data across sessions for train and testing and train the model with multiple sessions with dynamic numbers of subjects.
The first session of each session type in Table~\ref{tab:fumi-sessions} is used for testing, and all the others are used for training.
% The gaze interaction data for each session undergoes chronological partitioning, with 70\% allocated for training, 15\% for validation, and another 15\% for testing. 
% Specifically, the initial 70\% of gaze interactions constitute the training set, followed by the subsequent 15\% for validation, and the final 15\% for testing.
% \textbf{Evaluation method and metrics.}To validate our approach, we conducted examinations in both \textbf{transductive and inductive settings}. In the transductive task, we focused on predicting future links among observed gaze interactions during training. Conversely, in the inductive tasks, the aim was to predict future links for previously unobserved gaze interactions. Evaluation metrics included average precision accuracy (AP), Receiver Operating Characteristic Area Under the Curve (ROC-AUC), and f1 score. 
We use the Binary Cross Entropy loss and the Adam optimizer with the learning rate set to 0.001 with 1e-4 weight decay. 
The dataset is sampled to be 1 Hz, and batch size is set to match 10 sec time window, i.e., $10\times\text{(number of subjects)}$.
Furthermore, early stopping is incorporated with a patience parameter set to 10.

% \begin{algorithm} 
% \caption{Negative sampling strategy}\label{alg:Negative sampling}
% \begin{algorithmic} [ht]

% \State $Subjects \gets \{V_i,V_j,..., V_n\}$ \Comment{All the subjects of the sessions}
% \State $source \gets V_i$    \Comment{The participant gazing} 
% \State $destination \gets V_j$ \Comment{The participant being gazed}
% \State $possible\_negatives$: $[]$ \\
% \hfill\Comment{Ensuring subjects chosen is among the session}
%  \For{$V_{neg}$ in $Subjects$}
%     \If{$V_{neg}$ $\neq$ $source$} \Comment{Filter self gaze}
%         \If{$V_{neg}$ $\neq$ $destination$} \Comment{Filter positive sample}
%             \State $possible\_negatives$$ \gets V_{neg}$ 
%         \EndIf
%     \EndIf
%   \EndFor
% \end{algorithmic}
% \end{algorithm}
\section{Results and discussion}\label{results}
This section presents the outcomes derived from experimentation across various sessions. 
The results encompass both inductive and transductive studies. Furthermore, several ablation tests are included to facilitate the analysis of the impacts of batch sizes and the significance of gaze interaction features. 

\subsection{Performance on Next Gaze and Next Speaker Prediction}
\begin{table}[b]
\centering
\setlength\tabcolsep{7pt}
\begin{tabular}{@{}c|clcl|clcl@{}}
\toprule
                 & \multicolumn{4}{c|}{\textbf{Next Gaze}}                      & \multicolumn{4}{c}{\textbf{Next Speaker}}                   \\ 
                 % \midrule
                 & \multicolumn{2}{c}{F1-score} & \multicolumn{2}{c|}{Accuracy} & \multicolumn{2}{c}{F1-score} & \multicolumn{2}{c}{Accuracy} \\ \midrule
\textbf{History Model} & \multicolumn{2}{c}{30.1\%}   & \multicolumn{2}{c|}{59.6\%}   & \multicolumn{2}{c}{16.4\%}   & \multicolumn{2}{c}{83.8\%}   \\ 
\textbf{TGN}    & \multicolumn{2}{c}{67.1\%}   & \multicolumn{2}{c|}{83.8\%}   & \multicolumn{2}{c}{45.4\%}   & \multicolumn{2}{c}{86.7\%}   \\ \midrule
 &
  \multicolumn{2}{c}{{\color[HTML]{3531FF} +37.0\%}} &
  \multicolumn{2}{c|}{{\color[HTML]{3531FF} +24.2\%}} &
  \multicolumn{2}{c}{{\color[HTML]{3531FF} +29.0\%}} &
  \multicolumn{2}{c}{{\color[HTML]{3531FF} +3.0\%}} \\ \bottomrule
\end{tabular}
% \begin{tabular}{@{}c|clcl|clcl@{}}
% \toprule
%                  & \multicolumn{4}{c|}{\textbf{Next Gaze}}                      & \multicolumn{4}{c}{\textbf{Next Speaker}}                   \\ \midrule
%                  & \multicolumn{2}{c}{F1-score} & \multicolumn{2}{c|}{Accuracy} & \multicolumn{2}{c}{F1-score} & \multicolumn{2}{c}{Accuracy} \\ \midrule
% \textbf{History} & 30.1\%       ($\pm$11.3\%)      & 59.6\%        ($\pm$7.0\%)       & 16.4\%       ($\pm$6.5\%)       & 83.8\%       ($\pm$5.7\%)       \\ \midrule
% \textbf{Ours}    & 67.1\%       ($\pm$16.2\%)      & 83.8\%        ($\pm$4.2\%)       & 45.4\%       ($\pm$25.3\%)      & 86.7\%       ($\pm$4.1\%)       \\ \midrule
%  &
%   {\color[HTML]{3531FF} +37.0\%} &
%   {\color[HTML]{3531FF} } &
%   {\color[HTML]{3531FF} +24.2\%} &
%   {\color[HTML]{3531FF} } &
%   {\color[HTML]{3531FF} +29.0\%} &
%   {\color[HTML]{3531FF} } &
%   {\color[HTML]{3531FF} +3.0\%} &
%    \\ \bottomrule
% \end{tabular}
\caption{Performance on Next Gaze and Next Speaker Prediction}
\label{tab:result}
\end{table}
\textbf{Baseline.} Similar to the memorization baseline proposed by Poursafaei et al.~\cite{poursafaei2022towards}, we use a \textit{history model} as a baseline.
This model is not a learned model but uses the most recently observed gaze or speaker as a prediction.
We use this model as a reasonable baseline because gazing and speaking status are continuous.

Table~\ref{tab:result} presents the predictions on the next gaze (task A) and next speaker (task B).
% For the next speaker prediction, we further trained the graph model from the next gaze prediction.
Our approach significantly outperforms the baseline across all metrics for both tasks.
For the next gaze prediction, TGN achieved an improvement of 37.0\% in the F1-score and 24.2\% in accuracy over the history model, indicating its robustness and showing that the graph uses the features to predict the existence and non-existence of edges.
Similarly, in determining the next speaking status of each subjects, TGN marked a substantial increase of 29.0\% in F1-score and 3.0\% in accuracy.

Still, the F1-score of the next speaker is under 50\%, which could be due to a lack of information on features we used in this work.
Specifically, the average performance of D1 and D2 showed the lowest among all the sessions. 
D1 and D2 do not have facilitators in the group, thus all the subjects have the same role type (i.e., \textit{student}).
That is, the role type does not provide meaningful information to predict each subject's behavior in sessions D1 and D2.
Also, the subjects would behave differently than other sessions with facilitators.
However, the information we provided to the model barely includes the context to predict the speaking status.
We expect that adding information related to speaking status, such as body language, pitch, and speed of speaking, will give more context and help predict the next speaker better.

\subsection{Ablation Study}
To conduct a more in-depth analysis of our model and the representation of interaction features, we execute the following ablation study.
Initially, we compare two types of message encodings suggested in Sec.~\ref{sec:method-model-message}. 
Then, we replicate the ablation study conducted in the TGN paper for our data to compare different TGN variants as well as temporal graph model baselines. 
% Subsequently, we assess the significance of the features associated with gaze interaction.

\subsubsection{Message Encodings}\label{sec:result-message}
We compare the generalizability of the two encodings, BERT and one-hot, by comparing the performance as shown in Table~\ref{tab:result-abl-msg}.
One session (S14) with the \textit{Music Teacher} facilitator is used to train the model, and we report the average performance of sessions in each facilitator type.
Overall, the performance was slightly increased when using one-hot encoding, but mostly showed similar performance.
Mainly, a performance drop occurred in \textit{Musician} sessions, where different types of facilitators participated, while the most significant performance increase was in sessions without a facilitator (\textit{None}).

These performance differences could be explained by the difference between BERT and the one hot encoding approach: BERT is likely to map three types of facilitators to similar space, musician-music teacher and music teacher and teacher are closely related pairs using its language knowledge. 
However, the one-hot encoder encodes four roles to orthogonal space, and the relations need to be learned during training, which is not done with a single session of training. 
It is highly likely that the results are primarily relying on the performance of student's data points.
Thus, if we use features that would benefit from prior knowledge of language models, BERT messages could give the extra information.
On the other hand, one-hot messages allow learning from scratch, which has the potential to be less biased but focus more on the given dataset and task. 
In our case, one-hot encoding was chosen because the relations of different types of facilitators were available and could be learned through multi-session training while requiring a smaller number of parameters due to smaller message size (BERT \textit{768} vs. one-hot \textit{14}).
\begin{table}[ht]
\centering
\setlength\tabcolsep{5pt}
\begin{tabular}{@{}cccc@{}}
\toprule
\textbf{Facilitator Type} & \textbf{BERT}  & \textbf{One-hot} & \textit{\textbf{diff}}                 \\ \midrule
None                 & 70.9\%  ($\pm$3.7\%) & 76.7\% ($\pm$3.0\%)    & {\color[HTML]{3531FF} \textit{\textbf{+5.7\%}}} \\
Teacher              & 81.3\%  ($\pm$2.6\%) & 81.7\% ($\pm$2.7\%)    & \textit{+0.4\%}                        \\
Music Teacher        & 89.8\%  ($\pm$1.6\%) & 89.8\% ($\pm$1.3\%)    & \textit{+0.1\%}                        \\
Musician             & 89.0\% ($\pm$2.7\%)  & 86.6\% ($\pm$6.2\%)    & {\color[HTML]{CB0000} \textit{\textbf{-2.4\%}}} \\ \bottomrule
\end{tabular}
\caption{Comparison on Message Types for generalizability across different sessions: Model is trained with one session with Music Teacher Facilitator and Tested on all the other sessions, for the Next Gaze prediction}
\label{tab:result-abl-msg}
\end{table}

\subsubsection{TGN variants}\label{sec:result-tgnvariant}
\begin{figure}[ht] 
  \centering
  \includegraphics[width=\linewidth]{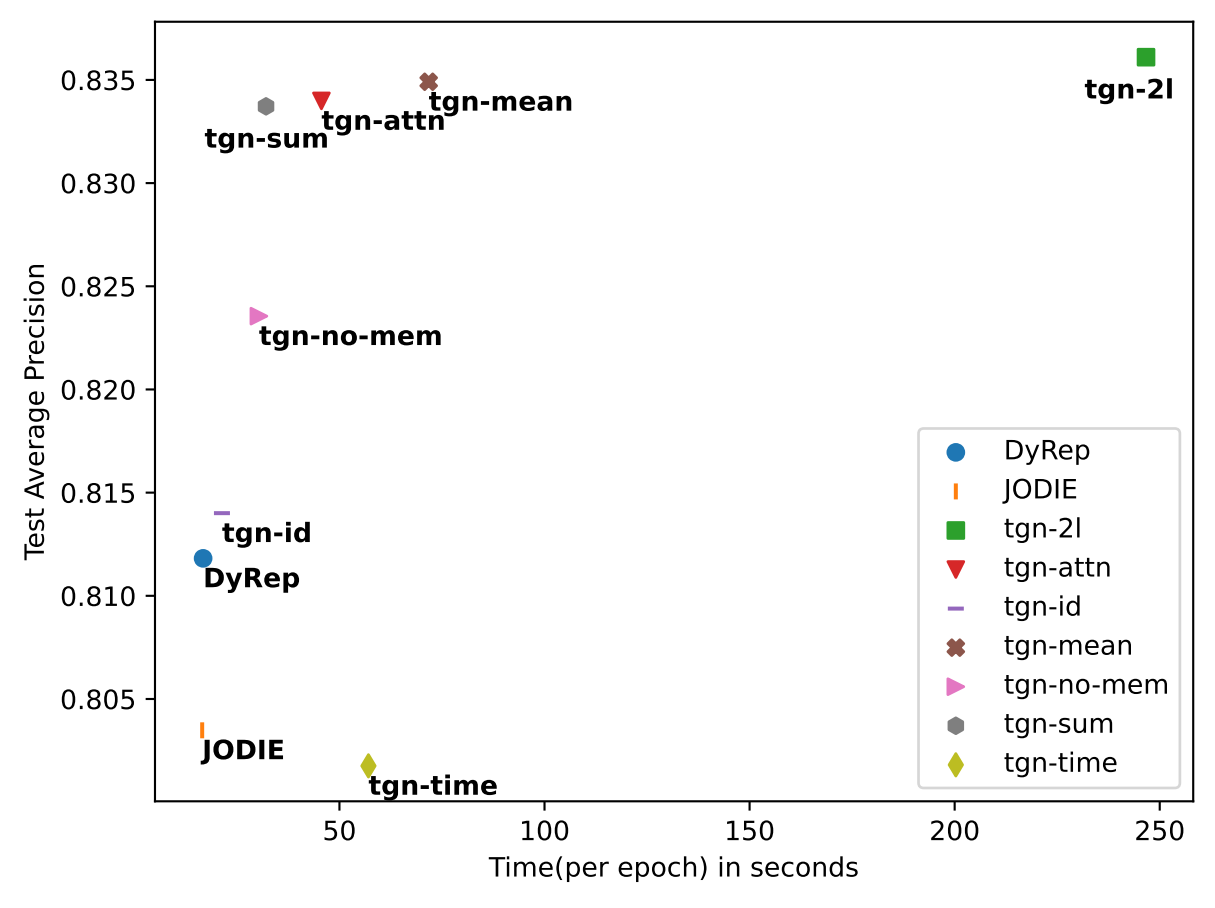}
  \caption{Comparison of different baselines based on Speed vs. Accuracy trade-off with model trained on S14.}\label{fig:TGN variants ablation}
  % \Description{This figure shows the comparison of different variants of TGN and 2 baseline methods for dynamic graphs, JODIE and DyRep. The comparison is done based on speed and accuracy. Overall TGN-attn variant performs the best. It is the same variant we have used for all our tests}
\end{figure}

We conduct a comparative analysis of outcomes achieved by various TGN variants and two baseline models of temporal graphs, namely \texttt{DyRep}~\cite{trivedi2019dyrep} and \texttt{Jodie}~\cite{kumar2019predicting}. 
The TGN variants \cite{rossi2020temporal} encompass diverse arrangements of modules governing memory embedding (with memory: \texttt{TGN-attn} vs. without memory: \texttt{-no-mem}), memory embedding module (identity: \texttt{TGN-id}, sum: \texttt{-sum}, time: \texttt{-time}, attention: \texttt{-attn}), message aggregator (last: \texttt{TGN-attn} vs. mean: \texttt{-mean}), and number of layers (1-layer: \texttt{TGN-attn} vs. 2-layers: \texttt{-2l}). 
As depicted in Figure \ref{fig:TGN variants ablation}, our findings suggest that the TGN-attn and TGN-mean variants offer the most favorable balance between accuracy and speed for modeling group interaction.

% \subsubsection{Significance of features}\hfill\\
% We further assess the model by exclusively training it on prior gaze interactions, excluding additional features such as speech, relative seating position, and role information. Exploring different batch sizes reveals that as batch size increases, the f1 score decreases. This is expected as explained in section 2.4, the batch size has large effect on accuracy due to memory degradation. However, this effect is more pronounced in the model without features, indicating its reliance on memorizing recent interactions. This supports our claim about the crucial role of features in comprehending and predicting gaze interactions. The comparison of average f1 scores for session 12 on all other sessions' data with and without features is depicted in Figure \ref{fig:Feature vs no features}.       

% \begin{figure}[!htbp] 
%   \centering
%   \includegraphics[width=\linewidth]{figures/feat_vs_nofeat_avg_f1score.pdf}
%   \caption{Effect of batch size with Features vs without using any features using model of Session No. 12}\label{fig:Feature vs no features}
%   % \Description{This figure depicts the effect of different batch sizes (100,200,500 and 1000) on f1 score by using model with features and model without features. From evaluation we can see that model without features performs significantly worse for understanding long term dependency compared to model with features.}
% \end{figure}

\section{Conclusion}
In this paper, we represented the dynamics of social interaction via temporal graph neural networks, using multi-modal features and their temporal dependencies to capture long-term interaction dynamics.  
% with the help of speech and environmental cues and showed the pivotal role they play in representing long-term interactions. 
Leveraging the generalization capabilities inherent in graph neural network models, we devised a model adaptable to varying subject counts. Capitalizing on the temporal nature of social interactions, we incorporated a temporal graph, integrating memory functionality. 

Even with a database containing diverse subjects and experimental conditions, we believe it's important to explore data from groups engaged in a wider variety of tasks that could serve as future work. The improvement of the TGN model by exploring different variations of message and memory modules could provide more insights into optimal ways to integrate the features of interaction.      

The proposed approach is relatively straightforward and allows for seamless extensions to include additional verbal elements (e.g., language, context, etc.), non-verbal features (e.g., actions, gestures, para-language, etc.), or human states, thereby serving as a foundation for future research.

% \begin{table} [H]
% \caption{Overview of Sessions}
%   \label{tab:table1}
%   \begin{tabular}{p{0.2\linewidth} p{0.5\linewidth} p{0.2\linewidth}}
%     \toprule
%     Session Number&Number of gaze interactions (respectively)&Number of Subjects present\\
%     \midrule
%     6,7,8 &115737,74818,73643&3\\
%     3,4,5,14,15,16, 17,18 &226647,1163840,149506,120469,112274,133109, 258631,236868&4\\
%     1,2,10,12 &155754, 221330, 319862, 265431 & 5 \\
%     9,11,13,19,20,21&369380,260031,281988,338761,214107,180404&6\\
%     \hline
%   \bottomrule
% \end{tabular}
% \end{table}

%                                   % the textheight of the last page by a suitable amount.
%                                   % This command does not take effect until the next page
%                                   % so it should come on the page before the last. Make
%                                   % sure that you do not shorten the textheight too much.

%%%%%%%%%%%%%%%%%%%%%%%%%%%%%%%%%%%%%%%%%%%%%%%%%%%%%%%%%%%%%%%%%%%%%%%%%%%%%%%%

%%%%%%%%%%%%%%%%%%%%%%%%%%%%%%%%%%%%%%%%%%%%%%%%%%%%%%%%%%%%%%%%%%%%%%%%%%%%%%%%



\section*{ACKNOWLEDGMENT}

The reported research is the results of a tripartite joint research collaboration between The University of Tokyo, Hiroo Gakuen Junior Senior High School, and Honda Research Institute Japan. The authors would like to thank the collaborators for their support and assistance in facilitating the implementation of experiments.

%%%%%%%%%%%%%%%%%%%%%%%%%%%%%%%%%%%%%%%%%%%%%%%%%%%%%%%%%%%%%%%%%%%%%%%%%%%%%%%%

\bibliographystyle{IEEEtran}
\bibliography{bib}

\end{document}